\documentclass[twocolumn,showpacs,preprintnumbers,amsmath,amssymb,amsfonts,superscriptaddress]{revtex4}

\usepackage{graphicx}
\usepackage{color}
\usepackage{ulem}

\newcommand{\be}{\begin{equation}}
\newcommand{\beq}{\begin{equation}}
\newcommand{\en}{\end{equation}}
\newcommand{\eeq}{\end{equation}}
\newcommand{\bea}{\begin{eqnarray}}
\newcommand{\ena}{\end{eqnarray}}
\newcommand{\hbo}{\hbox to 1 true cm {\hfill } }

\newcommand{\Os}{\mathcal{O}_{\rm S}}

\begin{document}

\preprint{KEK Preprint 2013-68, LLNL-JRNL-651959}

\title{Light composite scalar in eight-flavor QCD on the lattice}

\author{Yasumichi~Aoki}
\author{Tatsumi~Aoyama}
\author{Masafumi~Kurachi}
\author{Toshihide~Maskawa}
\author{Kohtaroh~Miura}
\author{Kei-ichi~Nagai}
\author{Hiroshi~Ohki}
\affiliation{Kobayashi-Maskawa Institute for the Origin of Particles and the Universe (KMI), Nagoya University, Nagoya, 464-8602, Japan}
\author{Enrico~Rinaldi}
\affiliation{Lawrence Livermore National Laboratory, Livermore, California 94550, USA}
\author{Akihiro~Shibata}
\affiliation{Computing Research Center, High Energy Accelerator Research Organization (KEK), Tsukuba 305-0801, Japan}
\author{Koichi~Yamawaki}
\author{Takeshi~Yamazaki}
\affiliation{Kobayashi-Maskawa Institute for the Origin of Particles and the Universe (KMI), Nagoya University, Nagoya, 464-8602, Japan}
\collaboration{LatKMI collaboration}
\noaffiliation

\date{
\today
}

\begin{abstract}

We present the first observation of a flavor-singlet scalar meson as light as the pion in $N_f=8$ QCD on the lattice, using the Highly Improved Staggered Quark action. 
Such a light scalar meson can be regarded as a composite Higgs with mass 125 GeV.
In accord with our previous lattice results showing that the theory exhibits walking behavior,
the light scalar may be a technidilaton, a pseudo Nambu-Goldstone boson of the approximate scale symmetry in walking technicolor.

\end{abstract}


\maketitle


Recently, a Higgs boson with mass around 125 GeV has been discovered 
at the Large Hadron Collider (LHC)~\cite{:2012gk,:2012gu}. 
While the current LHC data show good agreement with
the Standard model Higgs boson, 
there exists a possibility that the Higgs boson is a composite particle 
in an underlying strongly coupled gauge theory. 
A typical example is the walking technicolor theory, 
featuring approximate scale invariance and a large anomalous dimension, 
$\gamma_m \approx 1$~\cite{Yamawaki:1985zg} 
(see also similar works
~\cite{Holdom:1984sk, Akiba:1985rr,Appelquist:1986an}). 
Such a theory predicts a light composite Higgs, 
``technidilaton''~\cite{Yamawaki:1985zg},
emerging as a pseudo Nambu-Goldstone (NG) boson of 
the spontaneously broken approximate scale symmetry. 
It was shown
~\cite{Matsuzaki:2012xx,Matsuzaki:2012mk}
that the technidilaton is phenomenologically consistent 
with the current LHC data.

Thus, the most urgent theoretical task to test walking technicolor theories
would be to check whether or not such a light 
flavor-singlet scalar bound state 
exists from first-principle calculations with lattice gauge theory. 
Since the composite Higgs should be associated with the electroweak symmetry breaking, it must be predominantly a bound state of technifermions carrying 
electroweak charges, but not of technigluons having no electroweak charges (up to some mixings between them). Thus we look for a light flavor-singlet scalar 
meson in the correlator of fermionic operators on the lattice.

One of the most popular candidates for
walking technicolor theories is
QCD with a large number of (massless) flavors ($N_f$) 
in the fundamental representation. 
For the past few years, we have studied the SU(3) gauge theory with 
$N_f = 4, 8, 12$, and 16, in a common lattice setup~\cite{Aoki:2012eq,Aoki:2013xza,Aoki:2013zsa}.
(For reviews of lattice studies in search for candidates for walking technicolor theories, 
see~\cite{DelDebbio:2011rc,Neil:2012cb,Giedt:2012hg,Kuti:2014zz}.)

In $N_f=12$ QCD
we actually observed~\cite{Aoki:2013pca,Aoki:2013zsa}
a flavor-singlet scalar meson $(\sigma)$
lighter than the ``pion" having the quantum numbers 
corresponding to the NG pion ($\pi$) in the broken phase.
(Recently a light flavor-singlet scalar meson consistent
with ours was also observed 
by another group~\cite{Fodor:2014pqa} 
using a different lattice action.)

We found~\cite{Aoki:2012eq} that $N_f=12$ QCD  is consistent with
a conformal theory.
If it is a conformal theory,  it should have no bound states (``unparticle") 
in the exact chiral limit, and hence a light bound state
can only be formed in the presence of a
fermion mass $m_f$ which explicitly (not spontaneously) breaks 
the scale/chiral/electroweak symmetry. 

Hence such a light scalar meson in $N_f=12$ QCD  would not be
a composite Higgs associated with the spontaneous symmetry breaking.
However, its presence strongly suggests that
a walking theory would have a similar light scalar meson as a composite Higgs associated with the spontaneous scale/chiral/electroweak symmetry breaking, since in the walking theory
the gauge coupling is similar to that of 
a conformal theory with the role of the explicit breaking mass $m_f$
replaced by the dynamically generated fermion mass, $m_D$, 
arising from the spontaneous chiral symmetry breaking.

In this Letter we indeed observe such a light flavor-singlet scalar fermionic bound state
$\sigma$ as light as $\pi$  in $N_f = 8$ QCD, 
which we found~\cite{Aoki:2013xza}
is a candidate for walking technicolor, with the spontaneous breaking of chiral symmetry and 
large anomalous dimension near unity.  
Thus it can be a candidate for the composite Higgs (technidilaton) with a 125 GeV mass.
The preliminary results of this work were already reported in Ref.~\cite{Aoki:2013qxa}.


We carry out simulations of SU(3) gauge theory with eight fundamental fermions using 
two degenerate staggered fermion species with bare fermion mass $m_f$, 
where each species has four fermion degrees of freedom, called tastes.
We use a tree-level Symanzik gauge action and the Highly Improved Staggered Quark (HISQ)~\cite{Follana:2006rc} 
action without the tadpole improvement or the mass correction in the Naik term~\cite{Bazavov:2011nk}. 
The flavor symmetry breaking of this action is highly suppressed in QCD~\cite{Bazavov:2011nk}.
It is also true in our $N_f = 8$ QCD simulations, where the breaking is almost negligible
in the meson masses~\cite{Aoki:2013xza}.
At the same bare coupling $\beta \equiv 6/g^2 = 3.8$ as in our previous work~\cite{Aoki:2013xza}, 
we calculate the mass of the flavor-singlet scalar ($m_\sigma$) at five fermion masses,
$m_f = 0.015, 0.02, 0.03, 0.04$, and $0.06$,
to investigate the $m_f$ dependence of $m_\sigma$.
We use four volumes of spatial extent $L=18, 24, 30$, and 36,
with fixed aspect ratio $T/L = 4/3$, to check for finite size effects on $m_\sigma$.
All the simulation parameters are tabulated in Table~\ref{tab:simulations}.
In this Letter all dimensionful quantities are expressed in lattice units. 

We generate between $6400$ and $100000$ trajectories depending on 
the simulation parameters with 
the standard hybrid Monte Carlo algorithm using the MILC code version 7~\cite{MilcCode} 
with some modifications to suit our needs, such as Hasenbusch mass preconditioning~\cite{Hasenbusch:2001ne} 
to reduce the computational cost. 
For the thermalization we discard more than 2000 trajectories.
In some parameters we make several Markov chain streams
to collect thermalized configurations more efficiently.
The total numbers of configurations and 
Markov chain streams 
are tabulated in Table~\ref{tab:simulations}.
For the measurement of the flavor-singlet scalar mass
we use interpolating operators of the fermionic bilinear 
with the appropriate quantum numbers, $J^{PC} = 0^{++}$.
In this measurement
we use the MILC code~\cite{MilcCode} and exploit
GPGPU acceleration thanks to the QUDA library~\cite{Clark:2009wm}.
The measurements are performed every 2 trajectories.
The vacuum-subtracted disconnected correlator 
has large statistical noise; however, it is essential to obtain $m_\sigma$.
For the noise reduction, as in the $N_f = 12$ QCD 
calculation~\cite{Aoki:2013zsa},
we utilize a method~\cite{Venkataraman:1997xi}
based on the axial Ward-Takahashi identity~\cite{McNeile:2012xh}, 
which has been 
employed in the literature~\cite{Venkataraman:1997xi,Gregory:2007ev,McNeile:2012xh,Jin:2012dw}.
We use $64$ random sources spread in spacetime and color spaces for this noise-reduction method. 
The statistical errors are estimated by the jackknife method, with a bin size of 200 trajectories to eliminate autocorrelation sufficiently.

 \begin{table}
  \begin{tabular}{c|c|r|l|l|l}
     $m_f$ & $L^3 \times T$ & \multicolumn{1}{|c}{$N_{\rm cf}$[$N_{\rm st}$]} 
     & \multicolumn{1}{|c|}{$m_\sigma$} & \multicolumn{1}{c}{$m_\pi$} & \multicolumn{1}{|c}{$F_\pi$}\\
     \hline
     0.015& $36^3\! \times\! 48$ &  3200[2] & 0.155(21)($^{\ 0}_{41}$) & 0.1861(4)$^*$ & 0.0503(2)$^*$ \\
     0.02 & $36^3\! \times\! 48$ &  5000[1] & 0.190(17)($^{39}_{\ 0}$) & 0.2205(3)$^*$ & 0.0585(1)$^*$ \\
     0.02 & $30^3\! \times\! 40$ &  8000[1] & 0.201(21)($^{\ 0}_{60}$) & 0.2227(9)     & 0.0578(2) \\
     0.03 & $30^3\! \times\! 40$ & 16500[1] & 0.282(27)($^{24}_{\ 0}$) & 0.2812(2)$^*$ & 0.07140(9)$^*$ \\
     0.03 & $24^3\! \times\! 32$ & 36000[2] & 0.276(15)($^{6}_{0}$)    & 0.2832(14)    & 0.0715(4) \\
     0.04 & $30^3\! \times\! 40$ & 12900[3] & 0.365(43)($^{17}_{\ 0}$) & 0.3349(3)$^*$ & 0.0826(1)$^*$ \\
     0.04 & $24^3\! \times\! 32$ & 50000[2] & 0.322(19)($^{8}_{0}$)    & 0.3353(7)     & 0.0823(2) \\
     0.04 & $18^3\! \times\! 24$ &  9000[1] & 0.228(30)($^{\ 0}_{16}$) & 0.3421(29)    & 0.0823(5) \\
     0.06 & $24^3\! \times\! 32$ & 18000[1] & 0.46(7)($^{12}_{\ 0}$)   & 0.4295(6) & 0.1012(3) \\
     0.06 & $18^3\! \times\! 24$ &  9000[1] & 0.386(77)($^{12}_{\ 0}$) & 0.4317(15)    & 0.0999(5) \\
   \end{tabular}
   \caption{\label{tab:simulations} Simulation parameters for $N_f=8$ QCD at $\beta=3.8$. 
$N_{\rm cf}$($N_{\rm st}$) is the total number of gauge configurations (Markov chain streams). 
The second error of $m_\sigma$ is a systematic error coming from the fit range. 
The values for $m_\pi$ and $F_\pi$ are from Ref.~\cite{Aoki:2013xza}, 
but the ones with ($^*$) have been updated.
   }
 \end{table}


Since we employ the same fermion bilinear operator as in $N_f = 12$ QCD~\cite{Aoki:2013zsa},
in this Letter we describe it briefly.
We use the local fermionic bilinear operator of the $({\bf 1} \otimes {\bf 1})$
staggered spin-taste structure defined as
\begin{equation}
  \label{eq:fermionic-scalar-op}
  \Os(t) = \sum_{i=1}^{2}\sum_{\vec{x}} \overline{\chi}_i(\vec{x},t) \chi_i(\vec{x},t),
\end{equation}
where the index $i$ runs through different staggered fermion species. 
The correlator of the operator is given by
the connected $C(t)$ and also vacuum-subtracted disconnected $D(t)$ correlators, 
$\langle \Os(t) \Os^\dag(0) \rangle = 2 D(t) - C(t)$, 
where the factor in front of $D(t)$ comes from the number of species. 
Due to the staggered fermion symmetry, at large time, 
the correlator has two contributions from $\sigma$
and also its parity partner, which is 
a flavor non-singlet (taste non-singlet but species-singlet) 
pseudoscalar
(${\pi_{\rm \overline{SC}}}$)
\begin{equation}
  \label{eq:fermionic-combined}
2 D(t) - C(t) = A_\sigma(t) + (-1)^t A_{\pi_{\rm \overline{SC}}}(t),
\end{equation}
where $A_H(t) = A_H (e^{-m_H t} + e^{-m_H (T-t)})$, with $m_H$ 
being the mass of state $H$.
Since $-C(t)$ can be regarded as a flavor non-singlet scalar correlator, 
it should have contributions from the non-singlet scalar ($a_0$), and
its staggered parity partner, which is another flavor non-singlet 
(taste non-singlet and species non-singlet) pseudoscalar ($\pi_{\rm SC}$).
When $t$ is large, we can write
\begin{equation}
  \label{eq:fermionic-connected}
 -C(t)  =  A_{a_0}(t) + (-1)^t A_{\pi_{\rm SC}}(t).
\end{equation}
From Eq.~\eqref{eq:fermionic-combined} and Eq.~\eqref{eq:fermionic-connected}, 
at large time $2D(t)$ can be written as 
\begin{equation}
  \label{eq:fermionic-disconnected-only}
  2 D(t)  =  A_{\sigma}(t) - A_{a_0}(t) + (-1)^t
(A_{\pi_{\rm SC}}(t) - A_{\pi_{\rm \overline{SC}}}(t)).
\end{equation}
If the flavor symmetry is exact, all the flavor non-singlet pseudoscalars,
$\pi_{\rm \overline{SC}}$, $\pi_{\rm SC}$, and also the NG $\pi$, are degenerate. 
Furthermore, in the flavor symmetric limit, 
their amplitudes in Eq.~\eqref{eq:fermionic-disconnected-only} also coincide,
so that $A_{\pi_{\rm SC}}(t) = A_{\pi_{\rm \overline{SC}}}(t)$ in this limit.

\begin{figure}[!t]
  \includegraphics*[height=6cm]{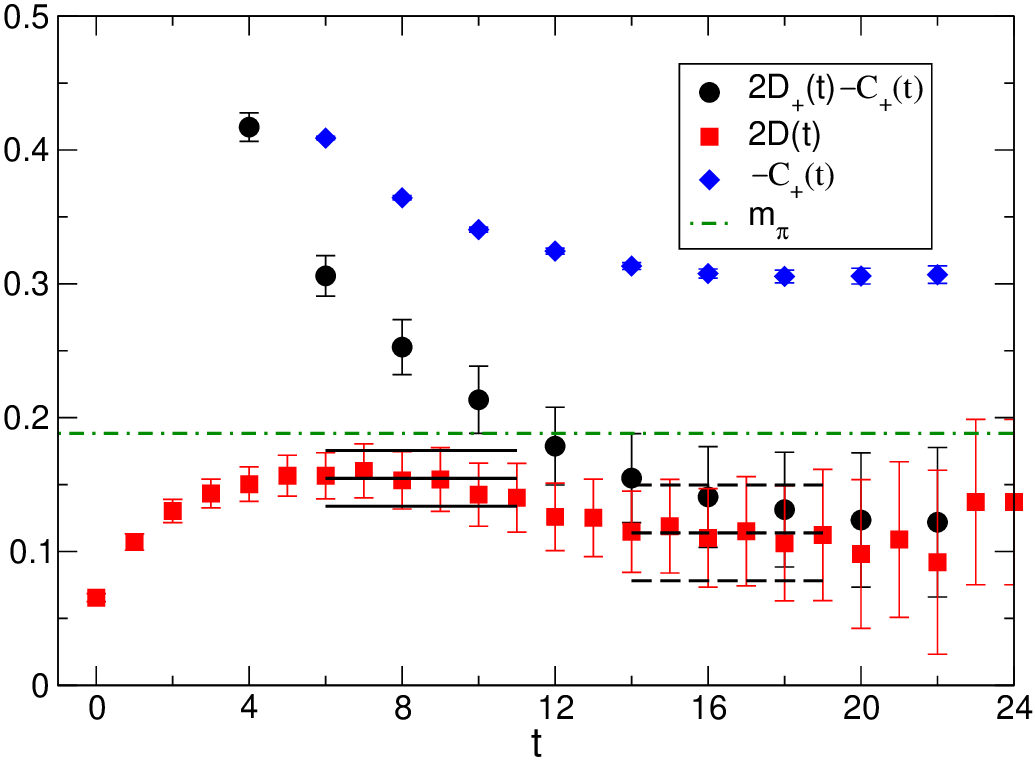}
  \caption{\label{fig:meff} Effective scalar mass $m_{\sigma}$ from correlators in Eq.~\eqref{eq:fermionic-combined}, 
  with the projection explained in the text, and in Eq.~\eqref{eq:fermionic-disconnected-only} for $L =36$ and $m_f=0.015$. 
  The solid and dashed lines highlight the fit results for $m_{\sigma}$ with statistical error band.
  The dashed-dotted line represents $m_\pi$. Effective mass of the projected connected correlator in Eq.~\eqref{eq:fermionic-connected}
is also plotted.}
\end{figure}


After applying the positive parity projection, $C_+(t) = 2C(t) + C(t+1) + C(t-1)$ at even $t$ to
minimize $A_{\pi_{\rm \overline{SC}}}(t)$ in Eq.~\eqref{eq:fermionic-combined},
we evaluate the effective mass of the projected correlator $2D_+(t)-C_+(t)$.
Figure~\ref{fig:meff} shows that the effective mass at large $t$ is almost equal to $m_\pi$, although the error is large. 
We also plot the effective mass of $2D(t)$ without the projection,
which does not have an oscillating behavior.
This means that the flavor symmetry breaking between $A_{\pi_{\rm SC}}(t)$ and 
$A_{\pi_{\rm \overline{SC}}}(t)$ in Eq.~\eqref{eq:fermionic-disconnected-only} is small.
The effective mass plateau of $2D(t)$ is statistically consistent with the one of $2D_+(t)-C_+(t)$ in the large time region.
Note that effective mass of $-C_+(t)$ is always 
larger than the one of $2D(t)$ in our simulations, as shown in Fig.~\ref{fig:meff}.
Since the plateau of $2D(t)$ appears at earlier time with smaller error than the one of $2D_+(t)-C_+(t)$,
we choose $2D(t)$ to extract $m_\sigma$ in all the parameters.
The earlier plateau suggests that the contribution of $a_0$ 
tends to cancel with that from excited flavor-singlet scalar states in $2D(t)$. 
It should be noted that, because of the small $m_\sigma$, comparable to $m_\pi$, 
the exponential damping of $D(t)$ is slow, 
and this helps preventing a rapid degradation of the signal-to-noise ratio.

We fit $2D(t)$ in the region $t=6$--11 by a single cosh form
assuming only $\sigma$ propagating in this region 
to obtain $m_\sigma$ for all the parameters. 
The fit result on $L=36$ at $m_f = 0.015$ is shown in Fig.~\ref{fig:meff}.
In this parameter it is possible to fit $2D(t)$ with a longer fit range,
while in some parameters the effective mass of $2D(t)$ in the large time region
is unstable with large error in the current statistics.
Thus, we choose this fixed fit range in all the parameters.
In order to estimate the systematic error coming from the fixed fit range,
we carry out another fit in a region at larger $t$ than the fixed one,
with the same number of data points. 
An example of this fit is shown in Fig.~\ref{fig:meff}.
We quote the difference between the two central values as the systematic error. 

\begin{figure}[t!]
  \includegraphics*[height=6cm]{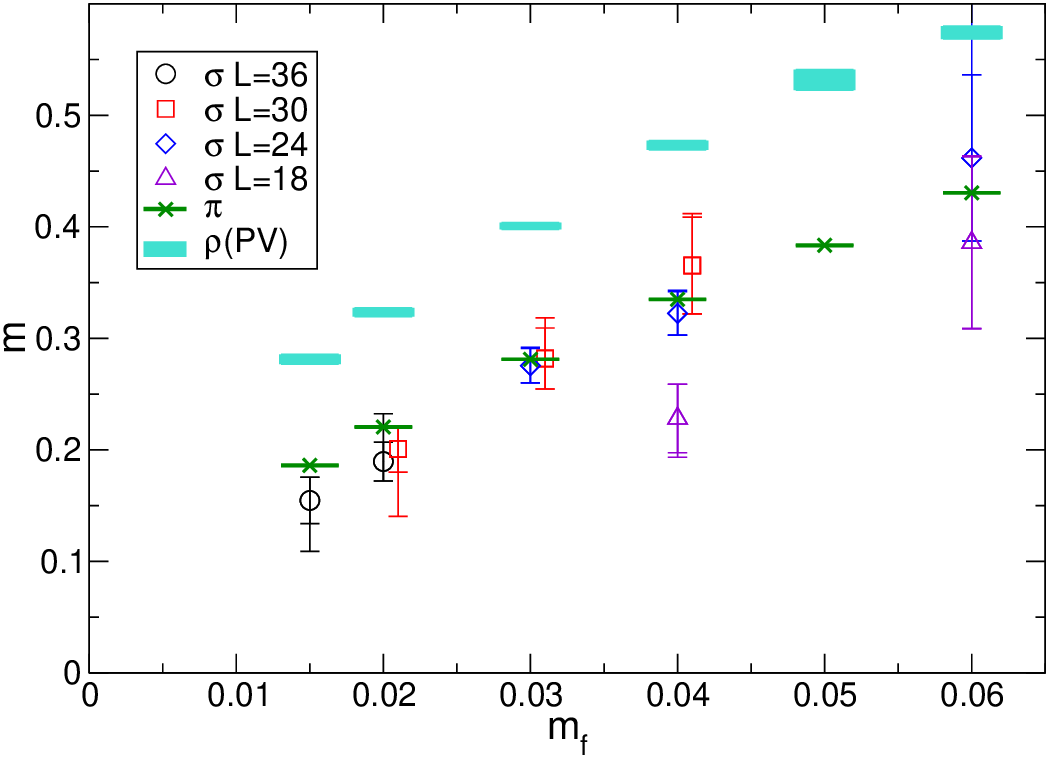}
  \caption{\label{fig:m_mf}  Mass of the flavor-singlet scalar $m_{\sigma}$ compared to the mass of NG pion $m_\pi$ 
  as a function of the fermion mass $m_f$. 
  Outer error represents the statistical and systematic uncertainties added in quadrature,
  while inner error is only statistical. 
  Square symbols are slightly shifted for clarity.
  Mass of vector meson with one standard deviation is expressed by full boxes.
  }
\end{figure}

The values of $m_\sigma$ and also $m_\pi$  
for all the parameters are summarized in Table~\ref{tab:simulations}. 
Figure~\ref{fig:m_mf} presents $m_\sigma$ as function of $m_f$ together with $m_\pi$.
These are our main results.
The data on the largest two volumes at each $m_f$, except for $m_f = 0.015$,
agree with each other, and suggest that 
finite size effects are negligible in our statistics. 
We find a clear signal that $\sigma$ is as light as $\pi$ for all the fermion masses we simulate.
This property is distinctly different from the one in usual QCD, where
$m_\sigma$ is clearly larger than $m_\pi$~\cite{Kunihiro:2003yj,Bernard:2007qf},
while it is similar to the one in $N_f = 12$ QCD observed in our previous study~\cite{Aoki:2013zsa}.
Thus, this might be regarded as a reflection of the approximate scale symmetry in this theory, no matter whether the main scale symmetry breaking in the far infrared comes from $m_f$ or $m_D$, as we noted before.
The figure also shows that our simulation region is far from heavy-fermion limit,
because the vector meson mass obtained from the $(\gamma_i\gamma_4 \otimes \xi_i\xi_4)$ operator,
denoted by $\rho$(PV), is clearly larger than $m_\pi$.

Although the accuracy of our data is not enough to make a clear conclusion for a chiral extrapolation, we shall report some results below.
While in the previous paper~\cite{Aoki:2013xza} we found that the data 
for $m_\pi$ and $F_\pi$, $\pi$ decay constant at each $m_f$,
are consistent with chiral perturbation theory (ChPT) in the region $m_f \leq 0.04$, 
the updated data~\cite{Aoki:2014xxx}, tabulated in Table~\ref{tab:simulations}, 
show consistency with ChPT 
in a somewhat smaller region $m_f \leq 0.03$.
Thus, we shall use the lightest three data with
the smallest error at each $m_f$, i.e., the two data on $L=36$ and the lightest data on $L=24$, 
in the following analyses.

\begin{figure}[t]
  \includegraphics*[height=6cm]{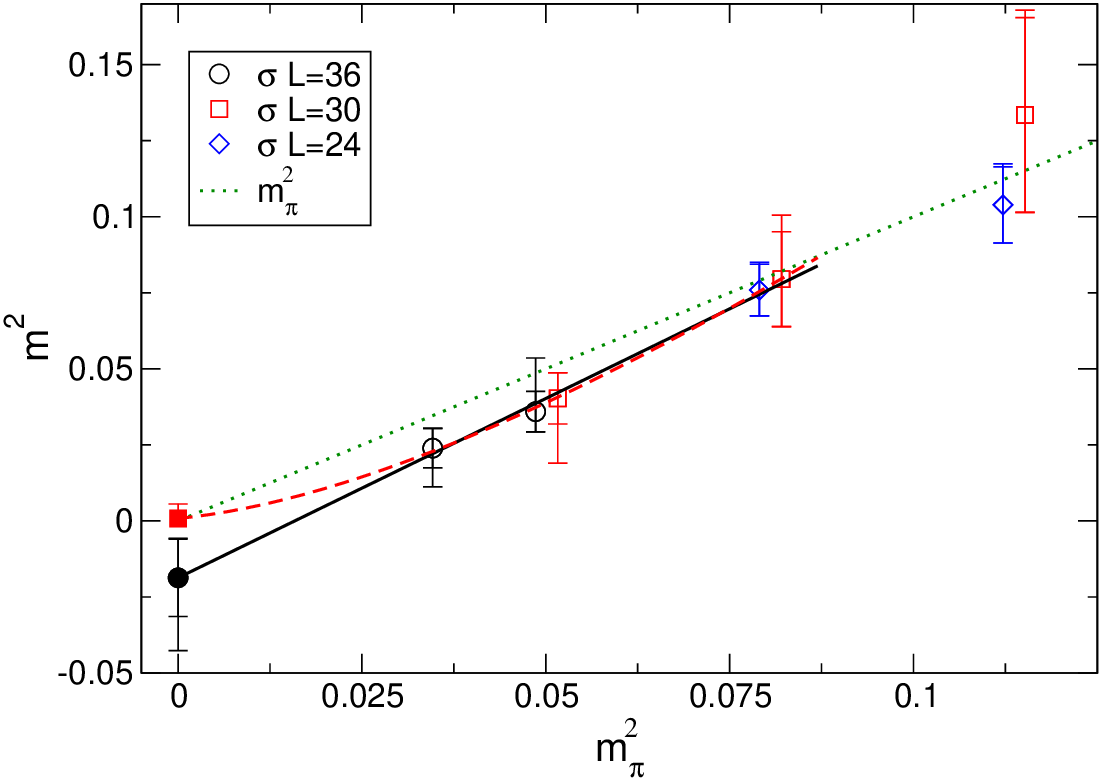}
  \caption{\label{fig:m2_mpi2} Mass squared of the flavor-singlet scalar 
  $m_{\sigma}^2$ as function of $m_\pi^2$.
  Outer error represents the statistical and systematic uncertainties added in quadrature,
  while inner error is only statistical.
  Open square symbols are slightly shifted for clarity.
  Result of a chiral extrapolation by the DChPT fit in Eq.~\eqref{eq:DChPT}
  is plotted by the solid line and full circle.
  Linear fit result, $m_\sigma = c_0 + c_1 m_f$, is also plotted by dashed curve and full square.
  Dotted line denotes $m_\sigma^2 = m_\pi^2$.
  }
\end{figure}

The validity of ChPT is intact even when the light $\sigma$ comparable with $\pi$ is involved in the chirally broken phase: 
the systematic power counting rule as a generalization of ChPT including 
$\sigma$ as a dilaton was established in Ref.~\cite{Matsuzaki:2013eva} 
(``dilaton ChPT (DChPT)") including computation of the chiral log effects.
At the leading order we have 
$m_\pi^2 = 4 m_f \langle \bar \psi \psi\rangle/F^2$ (Gell-Mann-Oakes-Renner relation) and 
\begin{equation}
m_\sigma^2 = d_0 + d_1 m_\pi^2\, ,
\label{eq:DChPT}
\end{equation}
where $d_0=m_\sigma^2|_{m_f=0} $ and
$d_1 = (3-\gamma_m)(1+\gamma_m)/4\cdot (N_f F^2)/F_\sigma^2$,
with $\gamma_m$ being mass anomalous dimension in the walking region, $F$ and $F_\sigma$ being the
decay constants of $\pi$ and $\sigma$, respectively, in the chiral limit. ($F/\sqrt{2}$ corresponds to 93 MeV for the usual QCD $\pi$.)
In the following fit, we ignore higher order terms including chiral log.
We plot $m_\sigma^2$ as a function of $m_\pi^2$ in Fig.~\ref{fig:m2_mpi2}.
The extrapolation to the chiral limit based on Eq.~\eqref{eq:DChPT} gives a reasonable $\chi^2/{\rm d.o.f.} = 0.27$, with a tiny value
in the chiral limit,
$d_0 = -0.019(13)(^{\ 3}_{20})$ where the first and second errors are statistical and systematic, respectively.
It agrees with zero with 1.4 standard deviation and shows a consistency with the NG nature of $\sigma$. 
The fit without the lightest point (with single volume)
gives a consistent result, showing that finite size effects
are not statistically relevant.
Although errors are large at this moment, it is very encouraging for obtaining a light technidilaton
to be identified with a composite Higgs with mass 125 GeV, with the value very close to  $F/\sqrt{2}\simeq 123\ {\rm GeV}$ of the one-family model with 4 weak-doublets, i.e., $N_f=8$. 
The value of $F$ from our data is estimated as
$F = 0.0202(13)(_{67}^{54})$, which
is updated from the previous paper~\cite{Aoki:2013xza}
using more statistics and a new smaller $m_f$ data. 
(If this scalar is to be identified with a composite Higgs, we expect $d_0 \sim F^2/2 \sim 0.0002$). 

From the value of $d_1$, we can read  $F_\sigma$, 
because the factor 
$(3-\gamma_m)(1+\gamma_m)/4$ is close to unity when we
use $\gamma_m = 0.6$--1.0~\cite{Aoki:2013xza}.
The value of  $F_\sigma$ is important to make a prediction of the couplings of 
the Higgs boson from the walking technicolor theory. 
The obtained slope is $d_1 = 1.18(24)(^{35}_{\ 7})$.
From $d_1$ we estimate $F_\sigma$ as $F_\sigma \sim \sqrt{N_f}F$,
in curious coincidence with the holographic estimate~\cite{Matsuzaki:2012xx}
and the linear sigma model.
Note that the property $d_1 \sim 1$ is another feature
different from usual QCD,
where a much larger slope was observed for $m_\pi > 670$ MeV~\cite{Kunihiro:2003yj}.

With our statistics we can also fit the data with an empirical form,
$m_\sigma = c_0 + c_1 m_f$, consistent with Eq.~\eqref{eq:DChPT} up to higher order corrections,
where we obtain $c_0 = 0.029(39)(^{\ 8}_{72})$ and the ratio $m_\sigma/(F/\sqrt{2}) = 2.0(2.7)(^{\ \ 8}_{5.1})$.
The fit result is plotted in Fig.~\ref{fig:m2_mpi2} as a function of
$m_\pi^2$ using a quadratic $m_f$ fit result for $m_\pi^2$.
Several other fits, such as a linear $m_\pi^2$ fit of $m_\sigma^2/F_\pi^2$,
are carried out, and they give reasonably consistent ratios with the one from $c_0$.
All the fit results suggest a possibility to reproduce the Higgs boson mass
within the large errors.

Note that due to the sizable error the $\sigma$ spectrum could also 
be consistent with the hyperscaling for the conformal theory. 
Different, more precisely measurable quantities are required to study 
if the theory is conformal or near-conformal~\cite{Aoki:2013xza,Aoki:2014xxx}.

We found that $N_f=8$ QCD behaves consistently with
a walking theory in the previous study~\cite{Aoki:2013xza}.
If our $\sigma$ is a candidate for the composite Higgs,
$m_\sigma$ should be non-zero in the chiral limit, and hence become larger than $m_\pi$ at $m_f$ smaller than 
the ones used in the current work. 
Note that it is predicted in Ref.~\cite{Matsuzaki:2013eva}
that chiral log effect of $\pi$ loops makes 
the $m_\pi^2$ dependence of $m_\sigma^2$
milder.
Therefore, observing $m_\sigma > m_\pi$ is an important future direction and 
is necessary to determine a precise value of $m_\sigma$ in the chiral limit,
though it requires more accurate data with a much smaller fermion mass.
Furthermore, in such a small $m_f$ region,
decay of $\sigma$ to two pions
should be taken into account to extract $m_\sigma$ using 
a variational method,
while $\sigma$ in this work cannot decay
due to the heavy fermion mass where $m_\sigma < 2 m_\pi$.
To check consistency of the ground state mass,
it is also important to calculate
$m_\sigma$ from gluonic operators as in 
our $N_f = 12$ QCD study~\cite{Aoki:2013pca,Aoki:2013zsa,Aoki:2013twa}.


In summary, 
using the same calculation techniques as 
in the study of $N_f = 12$ QCD~\cite{Aoki:2013zsa},
we have observed
clear signals of a flavor-singlet scalar as light as the pion
in $N_f = 8$ QCD, which was shown to be a candidate
for walking technicolor~\cite{Aoki:2013xza}.
Our simple chiral extrapolations suggest the possibility
of the existence of a very light flavor-singlet scalar to be identified 
with a composite Higgs, which may be the technidilaton, with mass 125 GeV,
although the errors on the extrapolated values are large.

Obviously, an important future direction is to obtain a more precise
value of $m_\sigma$ in the chiral limit 
to clarify whether this theory can really reproduce the Higgs boson mass of 125 GeV, and 
is really a candidate of theory beyond the standard model.
To do this, we should observe $m_\sigma > m_\pi$
discussed above, which could be regarded as another signal
of walking behavior.

\noindent {\it Acknowledgments.--}
Numerical simulation has been carried out on the supercomputer system $\varphi$ at KMI in Nagoya University, 
the computer facilities of the Information Technology Center in Nagoya University,
and the computer facilities of the Research Institute for Information Technology in Kyushu University. 
This work is supported by the JSPS Grant-in-Aid for Scientific Research (S) No.22224003, (C) No.23540300 (K.Y.), for Young Scientists (B) No.25800139 (H.O.) and No.25800138 (T.Y.), and also by Grants-in-Aid of the Japanese Ministry for Scientific Research on Innovative Areas No.23105708 (T.Y.). 
E. R. acknowledges the support of the U. S. Department of Energy under Contract DE-AC52-07NA27344 (LLNL).
We would like to thank Ed~Bennett, Philippe de~Forcrand, Anna~Hasenfratz, Julius~Kuti, Shinya~Matsuzaki,
and Volodya Miransky for fruitful discussions.

\bibliography{reference}

\begin{thebibliography}{32}
\expandafter\ifx\csname natexlab\endcsname\relax\def\natexlab#1{#1}\fi
\expandafter\ifx\csname bibnamefont\endcsname\relax
  \def\bibnamefont#1{#1}\fi
\expandafter\ifx\csname bibfnamefont\endcsname\relax
  \def\bibfnamefont#1{#1}\fi
\expandafter\ifx\csname citenamefont\endcsname\relax
  \def\citenamefont#1{#1}\fi
\expandafter\ifx\csname url\endcsname\relax
  \def\url#1{\texttt{#1}}\fi
\expandafter\ifx\csname urlprefix\endcsname\relax\def\urlprefix{URL }\fi
\providecommand{\bibinfo}[2]{#2}
\providecommand{\eprint}[2][]{\url{#2}}

\bibitem[{\citenamefont{ATLAS}(2012)}]{:2012gk}
\bibinfo{author}{\bibnamefont{ATLAS}} (\bibinfo{collaboration}{ATLAS
  Collaboration}), \bibinfo{journal}{Phys.Lett.}
  \textbf{\bibinfo{volume}{B716}}, \bibinfo{pages}{1} (\bibinfo{year}{2012}),
  \eprint{1207.7214}.

\bibitem[{\citenamefont{CMS}(2012)}]{:2012gu}
\bibinfo{author}{\bibnamefont{CMS}} (\bibinfo{collaboration}{CMS
  Collaboration}), \bibinfo{journal}{Phys.Lett.}
  \textbf{\bibinfo{volume}{B716}}, \bibinfo{pages}{30} (\bibinfo{year}{2012}),
  \eprint{1207.7235}.

\bibitem[{\citenamefont{Yamawaki et~al.}(1986)\citenamefont{Yamawaki, Bando,
  and Matumoto}}]{Yamawaki:1985zg}
\bibinfo{author}{\bibfnamefont{K.}~\bibnamefont{Yamawaki}},
  \bibinfo{author}{\bibfnamefont{M.}~\bibnamefont{Bando}}, \bibnamefont{and}
  \bibinfo{author}{\bibfnamefont{K.-i.} \bibnamefont{Matumoto}},
  \bibinfo{journal}{Phys.Rev.Lett.} \textbf{\bibinfo{volume}{56}},
  \bibinfo{pages}{1335} (\bibinfo{year}{1986}).

\bibitem[{\citenamefont{Holdom}(1985)}]{Holdom:1984sk}
\bibinfo{author}{\bibfnamefont{B.}~\bibnamefont{Holdom}},
  \bibinfo{journal}{Phys.Lett.} \textbf{\bibinfo{volume}{B150}},
  \bibinfo{pages}{301} (\bibinfo{year}{1985}).

\bibitem[{\citenamefont{Akiba and Yanagida}(1986)}]{Akiba:1985rr}
\bibinfo{author}{\bibfnamefont{T.}~\bibnamefont{Akiba}} \bibnamefont{and}
  \bibinfo{author}{\bibfnamefont{T.}~\bibnamefont{Yanagida}},
  \bibinfo{journal}{Phys.Lett.} \textbf{\bibinfo{volume}{B169}},
  \bibinfo{pages}{432} (\bibinfo{year}{1986}).

\bibitem[{\citenamefont{Appelquist et~al.}(1986)\citenamefont{Appelquist,
  Karabali, and Wijewardhana}}]{Appelquist:1986an}
\bibinfo{author}{\bibfnamefont{T.~W.} \bibnamefont{Appelquist}},
  \bibinfo{author}{\bibfnamefont{D.}~\bibnamefont{Karabali}}, \bibnamefont{and}
  \bibinfo{author}{\bibfnamefont{L.}~\bibnamefont{Wijewardhana}},
  \bibinfo{journal}{Phys.Rev.Lett.} \textbf{\bibinfo{volume}{57}},
  \bibinfo{pages}{957} (\bibinfo{year}{1986}).

\bibitem[{\citenamefont{Matsuzaki and Yamawaki}(2012)}]{Matsuzaki:2012xx}
\bibinfo{author}{\bibfnamefont{S.}~\bibnamefont{Matsuzaki}} \bibnamefont{and}
  \bibinfo{author}{\bibfnamefont{K.}~\bibnamefont{Yamawaki}},
  \bibinfo{journal}{Phys.Rev.} \textbf{\bibinfo{volume}{D86}},
  \bibinfo{pages}{115004} (\bibinfo{year}{2012}), \eprint{1209.2017}.

\bibitem[{\citenamefont{Matsuzaki and
  Yamawaki}(2013{\natexlab{a}})}]{Matsuzaki:2012mk}
\bibinfo{author}{\bibfnamefont{S.}~\bibnamefont{Matsuzaki}} \bibnamefont{and}
  \bibinfo{author}{\bibfnamefont{K.}~\bibnamefont{Yamawaki}},
  \bibinfo{journal}{Phys.Lett.} \textbf{\bibinfo{volume}{B719}},
  \bibinfo{pages}{378} (\bibinfo{year}{2013}{\natexlab{a}}),
  \eprint{1207.5911}.

\bibitem[{\citenamefont{Aoki et~al.}(2012)\citenamefont{Aoki, Aoyama, Kurachi,
  Maskawa, Nagai, Ohki, Shibata, Yamawaki, and Yamazaki}}]{Aoki:2012eq}
\bibinfo{author}{\bibfnamefont{Y.}~\bibnamefont{Aoki}},
  \bibinfo{author}{\bibfnamefont{T.}~\bibnamefont{Aoyama}},
  \bibinfo{author}{\bibfnamefont{M.}~\bibnamefont{Kurachi}},
  \bibinfo{author}{\bibfnamefont{T.}~\bibnamefont{Maskawa}},
  \bibinfo{author}{\bibfnamefont{K.-i.} \bibnamefont{Nagai}},
  \bibinfo{author}{\bibfnamefont{H.}~\bibnamefont{Ohki}},
  \bibinfo{author}{\bibfnamefont{A.}~\bibnamefont{Shibata}},
  \bibinfo{author}{\bibfnamefont{K.}~\bibnamefont{Yamawaki}}, \bibnamefont{and}
  \bibinfo{author}{\bibfnamefont{T.}~\bibnamefont{Yamazaki}}
  (\bibinfo{collaboration}{LatKMI Collaboration}), \bibinfo{journal}{Phys.Rev.}
  \textbf{\bibinfo{volume}{D86}}, \bibinfo{pages}{054506}
  (\bibinfo{year}{2012}), \eprint{1207.3060}.

\bibitem[{\citenamefont{Aoki et~al.}(2013{\natexlab{a}})\citenamefont{Aoki,
  Aoyama, Kurachi, Maskawa, Nagai, Ohki, Shibata, Yamawaki, and
  Yamazaki}}]{Aoki:2013xza}
\bibinfo{author}{\bibfnamefont{Y.}~\bibnamefont{Aoki}},
  \bibinfo{author}{\bibfnamefont{T.}~\bibnamefont{Aoyama}},
  \bibinfo{author}{\bibfnamefont{M.}~\bibnamefont{Kurachi}},
  \bibinfo{author}{\bibfnamefont{T.}~\bibnamefont{Maskawa}},
  \bibinfo{author}{\bibfnamefont{K.-i.} \bibnamefont{Nagai}},
  \bibinfo{author}{\bibfnamefont{H.}~\bibnamefont{Ohki}},
  \bibinfo{author}{\bibfnamefont{A.}~\bibnamefont{Shibata}},
  \bibinfo{author}{\bibfnamefont{K.}~\bibnamefont{Yamawaki}}, \bibnamefont{and}
  \bibinfo{author}{\bibfnamefont{T.}~\bibnamefont{Yamazaki}}
  (\bibinfo{collaboration}{LatKMI Collaboration}), \bibinfo{journal}{Phys.Rev.}
  \textbf{\bibinfo{volume}{D87}}, \bibinfo{pages}{094511}
  (\bibinfo{year}{2013}{\natexlab{a}}), \eprint{1302.6859}.

\bibitem[{\citenamefont{Aoki et~al.}(2013{\natexlab{b}})\citenamefont{Aoki,
  Aoyama, Kurachi, Maskawa, Nagai, Ohki, Rinaldi, Shibata, Yamawaki, and
  Yamazaki}}]{Aoki:2013zsa}
\bibinfo{author}{\bibfnamefont{Y.}~\bibnamefont{Aoki}},
  \bibinfo{author}{\bibfnamefont{T.}~\bibnamefont{Aoyama}},
  \bibinfo{author}{\bibfnamefont{M.}~\bibnamefont{Kurachi}},
  \bibinfo{author}{\bibfnamefont{T.}~\bibnamefont{Maskawa}},
  \bibinfo{author}{\bibfnamefont{K.-i.} \bibnamefont{Nagai}},
  \bibinfo{author}{\bibfnamefont{H.}~\bibnamefont{Ohki}},
  \bibinfo{author}{\bibfnamefont{E.}~\bibnamefont{Rinaldi}},
  \bibinfo{author}{\bibfnamefont{A.}~\bibnamefont{Shibata}},
  \bibinfo{author}{\bibfnamefont{K.}~\bibnamefont{Yamawaki}}, \bibnamefont{and}
  \bibinfo{author}{\bibfnamefont{T.}~\bibnamefont{Yamazaki}}
  (\bibinfo{collaboration}{LatKMI Collaboration}),
  \bibinfo{journal}{Phys.Rev.Lett.} \textbf{\bibinfo{volume}{111}},
  \bibinfo{pages}{162001} (\bibinfo{year}{2013}{\natexlab{b}}),
  \eprint{1305.6006}.

\bibitem[{\citenamefont{Del~Debbio}(2010)}]{DelDebbio:2011rc}
\bibinfo{author}{\bibfnamefont{L.}~\bibnamefont{Del~Debbio}},
  \bibinfo{journal}{PoS} \textbf{\bibinfo{volume}{LATTICE2010}},
  \bibinfo{pages}{004} (\bibinfo{year}{2010}), \eprint{1102.4066}.

\bibitem[{\citenamefont{Neil}(2011)}]{Neil:2012cb}
\bibinfo{author}{\bibfnamefont{E.~T.} \bibnamefont{Neil}},
  \bibinfo{journal}{PoS} \textbf{\bibinfo{volume}{LATTICE2011}},
  \bibinfo{pages}{009} (\bibinfo{year}{2011}), \eprint{1205.4706}.

\bibitem[{\citenamefont{Giedt}(2012)}]{Giedt:2012hg}
\bibinfo{author}{\bibfnamefont{J.}~\bibnamefont{Giedt}}, \bibinfo{journal}{PoS}
  \textbf{\bibinfo{volume}{LATTICE2012}}, \bibinfo{pages}{006}
  (\bibinfo{year}{2012}).

\bibitem[{\citenamefont{Kuti}(2013)}]{Kuti:2014zz}
\bibinfo{author}{\bibfnamefont{J.}~\bibnamefont{Kuti}}, \bibinfo{journal}{PoS}
  \textbf{\bibinfo{volume}{LATTICE2013}}, \bibinfo{pages}{004}
  (\bibinfo{year}{2013}).

\bibitem[{\citenamefont{Aoki et~al.}(2013{\natexlab{c}})\citenamefont{Aoki,
  Aoyama, Kurachi, Maskawa, Nagai, Ohki, Rinaldi, Shibata, Yamawaki, and
  Yamazaki}}]{Aoki:2013pca}
\bibinfo{author}{\bibfnamefont{Y.}~\bibnamefont{Aoki}},
  \bibinfo{author}{\bibfnamefont{T.}~\bibnamefont{Aoyama}},
  \bibinfo{author}{\bibfnamefont{M.}~\bibnamefont{Kurachi}},
  \bibinfo{author}{\bibfnamefont{T.}~\bibnamefont{Maskawa}},
  \bibinfo{author}{\bibfnamefont{K.-i.} \bibnamefont{Nagai}},
  \bibinfo{author}{\bibfnamefont{H.}~\bibnamefont{Ohki}},
  \bibinfo{author}{\bibfnamefont{E.}~\bibnamefont{Rinaldi}},
  \bibinfo{author}{\bibfnamefont{A.}~\bibnamefont{Shibata}},
  \bibinfo{author}{\bibfnamefont{K.}~\bibnamefont{Yamawaki}}, \bibnamefont{and}
  \bibinfo{author}{\bibfnamefont{T.}~\bibnamefont{Yamazaki}}
  (\bibinfo{collaboration}{LatKMI Collaboration})
  (\bibinfo{year}{2013}{\natexlab{c}}), \eprint{1302.4577}.

\bibitem[{\citenamefont{Fodor et~al.}(2014)\citenamefont{Fodor, Holland, Kuti,
  Nogradi, and Wong}}]{Fodor:2014pqa}
\bibinfo{author}{\bibfnamefont{Z.}~\bibnamefont{Fodor}},
  \bibinfo{author}{\bibfnamefont{K.}~\bibnamefont{Holland}},
  \bibinfo{author}{\bibfnamefont{J.}~\bibnamefont{Kuti}},
  \bibinfo{author}{\bibfnamefont{D.}~\bibnamefont{Nogradi}}, \bibnamefont{and}
  \bibinfo{author}{\bibfnamefont{C.~H.} \bibnamefont{Wong}},
  \bibinfo{journal}{PoS} \textbf{\bibinfo{volume}{LATTICE2013}},
  \bibinfo{pages}{062} (\bibinfo{year}{2014}), \eprint{1401.2176}.

\bibitem[{\citenamefont{Aoki et~al.}(2013{\natexlab{d}})\citenamefont{Aoki,
  Aoyama, Kurachi, Maskawa, Miura, Nagai, Ohki, Rinaldi, Shibata, Yamawaki
  et~al.}}]{Aoki:2013qxa}
\bibinfo{author}{\bibfnamefont{Y.}~\bibnamefont{Aoki}},
  \bibinfo{author}{\bibfnamefont{T.}~\bibnamefont{Aoyama}},
  \bibinfo{author}{\bibfnamefont{M.}~\bibnamefont{Kurachi}},
  \bibinfo{author}{\bibfnamefont{T.}~\bibnamefont{Maskawa}},
  \bibinfo{author}{\bibfnamefont{K.}~\bibnamefont{Miura}},
  \bibinfo{author}{\bibfnamefont{K.-i.} \bibnamefont{Nagai}},
  \bibinfo{author}{\bibfnamefont{H.}~\bibnamefont{Ohki}},
  \bibinfo{author}{\bibfnamefont{E.}~\bibnamefont{Rinaldi}},
  \bibinfo{author}{\bibfnamefont{A.}~\bibnamefont{Shibata}},
  \bibinfo{author}{\bibfnamefont{K.}~\bibnamefont{Yamawaki}}, \bibnamefont{and}
  \bibinfo{author}{\bibfnamefont{T.}~\bibnamefont{Yamazaki}}
  (\bibinfo{collaboration}{LatKMI Collaboration}),
  \bibinfo{journal}{PoS} \textbf{\bibinfo{volume}{LATTICE2013}},
  \bibinfo{pages}{070} (\bibinfo{year}{2013}{\natexlab{d}}),
  \eprint{1309.0711}.

\bibitem[{\citenamefont{Follana et~al.}(2007)}]{Follana:2006rc}
\bibinfo{author}{\bibfnamefont{E.}~\bibnamefont{Follana}} \bibnamefont{et~al.}
  (\bibinfo{collaboration}{HPQCD Collaboration, UKQCD Collaboration}),
  \bibinfo{journal}{Phys.Rev.} \textbf{\bibinfo{volume}{D75}},
  \bibinfo{pages}{054502} (\bibinfo{year}{2007}), \eprint{hep-lat/0610092}.

\bibitem[{\citenamefont{Bazavov et~al.}(2012)\citenamefont{Bazavov,
  Bhattacharya, Cheng, DeTar, Ding et~al.}}]{Bazavov:2011nk}
\bibinfo{author}{\bibfnamefont{A.}~\bibnamefont{Bazavov}},
  \bibinfo{author}{\bibfnamefont{T.}~\bibnamefont{Bhattacharya}},
  \bibinfo{author}{\bibfnamefont{M.}~\bibnamefont{Cheng}},
  \bibinfo{author}{\bibfnamefont{C.}~\bibnamefont{DeTar}},
  \bibinfo{author}{\bibfnamefont{H.}~\bibnamefont{Ding}}, \bibnamefont{et~al.},
  \bibinfo{journal}{Phys.Rev.} \textbf{\bibinfo{volume}{D85}},
  \bibinfo{pages}{054503} (\bibinfo{year}{2012}), \eprint{1111.1710}.

\bibitem[{Mil()}]{MilcCode}
\emph{\bibinfo{title}{Milc public lattice gauge theory code}},
  \urlprefix\url{http://physics.indiana.edu/~sg/milc.html}.

\bibitem[{\citenamefont{Hasenbusch}(2001)}]{Hasenbusch:2001ne}
\bibinfo{author}{\bibfnamefont{M.}~\bibnamefont{Hasenbusch}},
  \bibinfo{journal}{Phys.Lett.} \textbf{\bibinfo{volume}{B519}},
  \bibinfo{pages}{177} (\bibinfo{year}{2001}), \eprint{hep-lat/0107019}.

\bibitem[{\citenamefont{Clark et~al.}(2010)\citenamefont{Clark, Babich, Barros,
  Brower, and Rebbi}}]{Clark:2009wm}
\bibinfo{author}{\bibfnamefont{M.}~\bibnamefont{Clark}},
  \bibinfo{author}{\bibfnamefont{R.}~\bibnamefont{Babich}},
  \bibinfo{author}{\bibfnamefont{K.}~\bibnamefont{Barros}},
  \bibinfo{author}{\bibfnamefont{R.}~\bibnamefont{Brower}}, \bibnamefont{and}
  \bibinfo{author}{\bibfnamefont{C.}~\bibnamefont{Rebbi}},
  \bibinfo{journal}{Comput.Phys.Commun.} \textbf{\bibinfo{volume}{181}},
  \bibinfo{pages}{1517} (\bibinfo{year}{2010}), \eprint{0911.3191}.

\bibitem[{\citenamefont{Venkataraman and Kilcup}(1997)}]{Venkataraman:1997xi}
\bibinfo{author}{\bibfnamefont{L.}~\bibnamefont{Venkataraman}}
  \bibnamefont{and} \bibinfo{author}{\bibfnamefont{G.}~\bibnamefont{Kilcup}},
  \bibinfo{journal}{Phys.Rev.D}  (\bibinfo{year}{1997}),
  \eprint{hep-lat/9711006}.

\bibitem[{\citenamefont{McNeile et~al.}(2013)\citenamefont{McNeile, Bazavov,
  Davies, Dowdall, Hornbostel et~al.}}]{McNeile:2012xh}
\bibinfo{author}{\bibfnamefont{C.}~\bibnamefont{McNeile}},
  \bibinfo{author}{\bibfnamefont{A.}~\bibnamefont{Bazavov}},
  \bibinfo{author}{\bibfnamefont{C.}~\bibnamefont{Davies}},
  \bibinfo{author}{\bibfnamefont{R.}~\bibnamefont{Dowdall}},
  \bibinfo{author}{\bibfnamefont{K.}~\bibnamefont{Hornbostel}},
  \bibnamefont{et~al.}, \bibinfo{journal}{Phys.Rev.}
  \textbf{\bibinfo{volume}{D87}}, \bibinfo{pages}{034503}
  (\bibinfo{year}{2013}), \eprint{1211.6577}.

\bibitem[{\citenamefont{Gregory et~al.}(2008)\citenamefont{Gregory, Irving,
  Richards, and McNeile}}]{Gregory:2007ev}
\bibinfo{author}{\bibfnamefont{E.~B.} \bibnamefont{Gregory}},
  \bibinfo{author}{\bibfnamefont{A.~C.} \bibnamefont{Irving}},
  \bibinfo{author}{\bibfnamefont{C.~M.} \bibnamefont{Richards}},
  \bibnamefont{and} \bibinfo{author}{\bibfnamefont{C.}~\bibnamefont{McNeile}},
  \bibinfo{journal}{Phys.Rev.} \textbf{\bibinfo{volume}{D77}},
  \bibinfo{pages}{065019} (\bibinfo{year}{2008}), \eprint{0709.4224}.

\bibitem[{\citenamefont{Jin and Mawhinney}(2011)}]{Jin:2012dw}
\bibinfo{author}{\bibfnamefont{X.-Y.} \bibnamefont{Jin}} \bibnamefont{and}
  \bibinfo{author}{\bibfnamefont{R.~D.} \bibnamefont{Mawhinney}},
  \bibinfo{journal}{PoS} \textbf{\bibinfo{volume}{LATTICE2011}},
  \bibinfo{pages}{066} (\bibinfo{year}{2011}), \eprint{1203.5855}.

\bibitem[{\citenamefont{Kunihiro et~al.}(2004)}]{Kunihiro:2003yj}
\bibinfo{author}{\bibfnamefont{T.}~\bibnamefont{Kunihiro}} \bibnamefont{et~al.}
  (\bibinfo{collaboration}{SCALAR Collaboration}), \bibinfo{journal}{Phys.Rev.}
  \textbf{\bibinfo{volume}{D70}}, \bibinfo{pages}{034504}
  (\bibinfo{year}{2004}), \eprint{hep-ph/0310312}.

\bibitem[{\citenamefont{Bernard et~al.}(2007)\citenamefont{Bernard, DeTar, Fu,
  and Prelovsek}}]{Bernard:2007qf}
\bibinfo{author}{\bibfnamefont{C.}~\bibnamefont{Bernard}},
  \bibinfo{author}{\bibfnamefont{C.~E.} \bibnamefont{DeTar}},
  \bibinfo{author}{\bibfnamefont{Z.}~\bibnamefont{Fu}}, \bibnamefont{and}
  \bibinfo{author}{\bibfnamefont{S.}~\bibnamefont{Prelovsek}},
  \bibinfo{journal}{Phys.Rev.} \textbf{\bibinfo{volume}{D76}},
  \bibinfo{pages}{094504} (\bibinfo{year}{2007}), \eprint{0707.2402}.

\bibitem[{\citenamefont{Aoki et~al.}()\citenamefont{Aoki, Aoyama, Kurachi,
  Maskawa, Miura, Nagai, Ohki, Rinaldi, Shibata, Yamawaki
  et~al.}}]{Aoki:2014xxx}
\bibinfo{author}{\bibfnamefont{Y.}~\bibnamefont{Aoki}},
  \bibinfo{author}{\bibfnamefont{T.}~\bibnamefont{Aoyama}},
  \bibinfo{author}{\bibfnamefont{M.}~\bibnamefont{Kurachi}},
  \bibinfo{author}{\bibfnamefont{T.}~\bibnamefont{Maskawa}},
  \bibinfo{author}{\bibfnamefont{K.}~\bibnamefont{Miura}},
  \bibinfo{author}{\bibfnamefont{K.-i.} \bibnamefont{Nagai}},
  \bibinfo{author}{\bibfnamefont{H.}~\bibnamefont{Ohki}},
  \bibinfo{author}{\bibfnamefont{E.}~\bibnamefont{Rinaldi}},
  \bibinfo{author}{\bibfnamefont{A.}~\bibnamefont{Shibata}},
  \bibinfo{author}{\bibfnamefont{K.}~\bibnamefont{Yamawaki}}, \bibnamefont{and}
  \bibinfo{author}{\bibfnamefont{T.}~\bibnamefont{Yamazaki}}
  (\bibinfo{collaboration}{LatKMI Collaboration}),
  \bibinfo{note}{in preparation.}

\bibitem[{\citenamefont{Matsuzaki and
  Yamawaki}(2013{\natexlab{b}})}]{Matsuzaki:2013eva}
\bibinfo{author}{\bibfnamefont{S.}~\bibnamefont{Matsuzaki}} \bibnamefont{and}
  \bibinfo{author}{\bibfnamefont{K.}~\bibnamefont{Yamawaki}}
  (\bibinfo{year}{2013}{\natexlab{b}}), \eprint{1311.3784}.

\bibitem[{\citenamefont{Aoki et~al.}(2013{\natexlab{e}})\citenamefont{Aoki,
  Aoyama, Kurachi, Maskawa, Miura, Nagai, Ohki, Rinaldi, Shibata, Yamawaki
  et~al.}}]{Aoki:2013twa}
\bibinfo{author}{\bibfnamefont{Y.}~\bibnamefont{Aoki}},
  \bibinfo{author}{\bibfnamefont{T.}~\bibnamefont{Aoyama}},
  \bibinfo{author}{\bibfnamefont{M.}~\bibnamefont{Kurachi}},
  \bibinfo{author}{\bibfnamefont{T.}~\bibnamefont{Maskawa}},
  \bibinfo{author}{\bibfnamefont{K.}~\bibnamefont{Miura}},
  \bibinfo{author}{\bibfnamefont{K.-i.} \bibnamefont{Nagai}},
  \bibinfo{author}{\bibfnamefont{H.}~\bibnamefont{Ohki}},
  \bibinfo{author}{\bibfnamefont{E.}~\bibnamefont{Rinaldi}},
  \bibinfo{author}{\bibfnamefont{A.}~\bibnamefont{Shibata}},
  \bibinfo{author}{\bibfnamefont{K.}~\bibnamefont{Yamawaki}}, \bibnamefont{and}
  \bibinfo{author}{\bibfnamefont{T.}~\bibnamefont{Yamazaki}}
  (\bibinfo{collaboration}{LatKMI Collaboration}),
  \bibinfo{journal}{PoS} \textbf{\bibinfo{volume}{LATTICE2013}},
  \bibinfo{pages}{073} (\bibinfo{year}{2013}{\natexlab{e}}),
  \eprint{1310.0963}.

\end{thebibliography}

\end{document}